\documentclass[%
 aip,
 amsmath,amssymb,
 reprint,
]{revtex4-1}
\usepackage{bm}
\usepackage{amsmath}
\usepackage{amsfonts}
\usepackage{amssymb}
\usepackage{color}
\usepackage{dcolumn}
\usepackage[T1]{fontenc}
\usepackage{graphicx}
\usepackage[utf8]{inputenc}
\usepackage{mathptmx}
\usepackage{tabularx}
\usepackage{xfrac}
\newcommand{\PAGS}{PrAlGe$_{1-x}$Si$_{x}$}

\newcommand{\C}{$^\circ$C}
\newcommand{\angstrom}{\textup{\AA}}
\begin{document}

\preprint{AIP/123-QED}

\title[Invited Contribution]{Transition from Intrinsic to Extrinsic Anomalous Hall Effect in the Ferromagnetic Weyl Semimetal PrAlGe$_{1-x}$Si$_x$}

 \author{Hung-Yu Yang}
\affiliation{Department of Physics, Boston College, Chestnut Hill, MA 02467, USA}
\author{Bahadur~Singh}
\affiliation{Department of Physics, Northeastern University, Boston, MA 02115, USA}
\affiliation{SZU-NUS Collaborative Center and International Collaborative Laboratory of 2D Materials for Optoelectronic Science $\&$ Technology, Engineering Technology Research Center for 2D Materials Information Functional Devices and Systems of Guangdong Province, College of Optoelectronic Engineering, Shenzhen University, Shenzhen 518060, China}
\author{Baozhu~Lu}
\affiliation{Department of Physics and Temple Materials Institute, Temple University, USA}
\author{Cheng-Yi~Huang}
\affiliation{Institute of Physics, Academia Sinica, Taipei 11529, Taiwan}
\author{Faranak~Bahrami}
\affiliation{Department of Physics, Boston College, Chestnut Hill, MA 02467, USA}
\author{Wei-Chi~Chu}
\affiliation{Department of Physics, Northeastern University, Boston, MA 02115, USA}
\author{David~Graf}
\affiliation{National High Magnetic Field Laboratory, Tallahassee, FL 32310, USA}
\author{Shin-Ming~Huang}
\affiliation{Department of Physics, National Sun Yat-sen University, Kaohsiung 80424, Taiwan}
\author{Baokai Wang}
\affiliation{Department of Physics, Northeastern University, Boston, MA 02115, USA}
\author{Hsin Lin}
\affiliation{Institute of Physics, Academia Sinica, Taipei 11529, Taiwan}
\author{Darius~Torchinsky}
\affiliation{Department of Physics and Temple Materials Institute, Temple University, USA}
\author{Arun~Bansil}
\affiliation{Department of Physics, Northeastern University, Boston, MA 02115, USA}
\author{Fazel~Tafti}
\affiliation{Department of Physics, Boston College, Chestnut Hill, MA 02467, USA}
 \email{fazel.tafti@bc.edu}

\date{\today}

\begin{abstract}
Recent reports of a large anomalous Hall effect (AHE) in ferromagnetic Weyl semimetals (FM WSM) have led to a resurgence of interest in this enigmatic phenomenon.
However, due to a lack of tunable materials, the interplay between the intrinsic mechanism caused by Berry curvature and extrinsic mechanisms due to scattering remains unclear in FM WSMs.
In this contribution, we present a thorough investigation of both the extrinsic and intrinsic AHE in a new family of FM WSMs, \PAGS, where $x$ can be tuned continuously.
From DFT calculations, we show that the two end members, PrAlGe and PrAlSi, have different Fermi surfaces but similar Weyl node structures.
Experimentally, we observe moderate changes in the anomalous Hall coefficient ($R_S$) but significant changes in the ordinary Hall coefficient ($R_0$) in \PAGS\ as a function of $x$, confirming a change of Fermi surface.
By comparing the magnitude of $R_0$ and $R_S$, we identify two regimes; $|R_0|<|R_S|$ when $x\le0.5$ and $|R_0|>|R_S|$ when $x>0.5$.
Through a detailed scaling analysis, we discover a universal anomalous Hall conductivity (AHC) from intrinsic contribution when $x\le 0.5$.
Such universal AHC is absent when $x>0.5$.
Thus, we point out the significance of the extrinsic mechanisms in FM WSMs and report the first observation of a transition from intrinsic to extrinsic AHE in \PAGS.
\end{abstract}

\maketitle

\section{\label{intro} Introduction}
The Hall effect in ferromagnets is commonly characterized by the following empirical formula for the Hall resistivity $\rho_{xy}$~\cite{pugh_hall_1953,nagaosa_anomalous_2010}:
\begin{equation}
\label{hall}
\rho_{xy}=\rho_{xy}^O+\rho_{xy}^A=R_0H_z+ R_sM_z,
\end{equation}
where $R_0$ and $R_s$ are ordinary and anomalous Hall coefficients, respectively.
The ordinary Hall effect ($\rho_{xy}^O$) is linearly proportional to the applied field ($H_z$) through $R_0$ which is simply determined by carrier concentration in a single-band metal.
The anomalous Hall effect (AHE) $\rho_{xy}^A$, however, is proportional to the ferromagnetic moment $M_z$ through $R_S$ and can be due to a complicated combination of extrinsic and intrinsic mechanisms.
The main extrinsic mechanisms are skew-scattering~\cite{smit_spontaneous_1958} and side-jump~\cite{berger_side-jump_1970}, both of which are related to the scattering between electrons and impurities with spin-orbit coupling.
In contrast, the intrinsic (Karplus-Luttinger) mechanism originates from an anomalous velocity due to a phase shift in the electronic wave-packet which is independent of impurities~\cite{karplus_hall_1954,sundaram_wave-packet_1999}.
Since the reformulation of the intrinsic mechanism in terms of Berry curvature~\cite{sundaram_wave-packet_1999,jungwirth_anomalous_2002}, this concept has been successfully applied to explain the AHE in the canonical AHE material, bcc iron~\cite{yao_first_2004}.
The intrinsic mechanism is gaining increasingly more attention because it is also applicable to the AHE in Weyl semimetal (WSM) where the Weyl nodes, monopoles of Berry curvature, can potentially generate a large AHE\cite{burkov_anomalous_2014}.
Recently, several WSMs have been found to exhibit such large AHE, including pyrochlore iridates (Nd$_2$Ir$_2$O$_7$)~\cite{witczak-krempa_correlated_2014,ueda2018spontaneous}, Heusler and half-Heusler compounds (Co$_2$MnGa, GdPtBi)~\cite{belopolski_discovery_2019,manna_heusler_2018,suzuki2016large}, and ferromagnetic (FM) WSMs such as shandite structures (Co$_3$Sn$_2$S$_2$)~\cite{liu_giant_2018}.
All of these discoveries have been interpreted as intrinsic AHE and overlooked the importance of the extrinsic contribution.
For example, in the topological ferromagnet Fe$_3$Sn$_2$, the intrinsic contribution to AHE is confirmed via a scaling analysis, but the extrinsic contribution could be five times larger than the intrinsic one~\cite{ye2018massive,tian2009proper}.
Also, most studies of AHE are based on one single compound and therefore, can not distinguish the extrinsic from intrinsic contributions.
One experimental approach to address this issue would be to maintain the structure of the Weyl nodes but change the Fermi surface (or vice versa) across a series of compositions and find out the relative magnitude of extrinsic and intrinsic AHE.
This is precisely the subject of the present article that explores such a possibility in the FM WSMs PrAlGe$_{1-x}$Si$_x$.
We study the AHE in PrAlGe$_{1-x}$Si$_x$ with $x=0, 0.25, 0.5, 0.7, 0.85$ and $1$ to investigate both intrinsic and extrinsic contributions to the AHE in this tunable FM WSM family.
Although the two endpoints PrAlGe and PrAlSi are both FM WSMs with equal number of Weyl nodes, we reveal a transition of the AHE from an intrinsic regime ($x\leq 0.5$) to an extrinsic regime ($x>0.5$).
The significance of our results is two fold.
First, we demonstrate the importance of extrinsic contributions to AHE even in a FM WSM with robust Weyl nodes.
Second, we reveal a transition from intrinsic to extrinsic AHE in the same family of FM WSMs and show the possibility of tuning AHE in topological semimetals.

\section{\label{method} Experimental Methods}
\emph{Crystal growth}-- Single crystals of \PAGS\ were grown using a self-flux method from Pr ingots (99.00\%, Alfa Aesar), Al lumps (99.5\%, Alfa Aesar) Ge pieces (99.999\%+, Alfa Aesar) and Si lumps (99.999\%+, Alfa Aesar).
The starting chemicals were mixed with the mole ratio Pr:Al:Ge:Si = 1:10:$1-y$:$y$, placed in a crucible inside an evacuated quartz tube, heated to 1000~\C\ at 180~\C/hour, stayed at 1000~\C\ for 12 hours, cooled to 700~\C\ at 6~\C/hour, and annealed at 700~\C\ for another 12 hours.
Then, the tube was centrifuged to remove the excess Al flux.
All crystals of \PAGS\ were plate-like with the surface of the plate normal to the $c$-axis and its edges along the $a$-axis.
The chemical composition of each crystal was determined by energy dispersive X-ray spectroscopy (EDX) using a ZEISS Ultra-55 filed emission scanning electron microscope equipped with an EDAX detector.
Our EDX analysis in Appendix~\ref{app_EDX} shows that $x=y$ in \PAGS\ samples grown with $y=0, 0.5$, and $1$.
However, samples with $y=0.75$ and $0.9$, turn out to have $x=0.7$ and $0.85$, respectively (Table~\ref{T1} in Appendix~\ref{app_EDX}).
We note that in the literature, single crystals of PrAlGe made by flux growth and floating zone technique show slightly different properties\cite{meng_large_2019,puphal2019bulk}.
The resistivity and magnetization behavior of our PrAlGe samples behave similarly to the ones studied in Ref. \cite{meng_large_2019} with a residual resistivity ratio RRR~$\approx2.2$ that is $70\%$ larger than RRR~$\approx1.3$ in Ref. \cite{puphal2019bulk}.
Also, the EDX results in Appendix~\ref{app_EDX} show that our samples are not Al-rich unlike the samples in Ref.~\cite{puphal2019bulk} due to a smaller quantity of Al-flux used here.

\emph{Measurements}-- Electrical resistivity was measured with a standard four-probe technique and the heat capacity was measured with a relaxation time method in a Quantum Design physical property measurement system (PPMS) Dynacool.
The dc magnetization experiment was conducted on the vibrating sample magnetometer in a Quantum Design MPMS3.
The high-field experiment was performed in a $35$~T dc Bitter magnet inside a $^3$He fridge at a base temperature of $300$~mK, at the National High Magnetic Field laboratory in Tallahassee.
All samples used for transport measurements were carefully sanded to remove the residual Al-flux and to have the ideal bar geometry for the determination of resistivity.

\emph{Calculations}-- Electronic structure calculations were performed within the framework of density functional theory (DFT) using the experimental lattice parameters and a projector augmented wave (PAW) method implemented in the Vienna ab-initio simulation package (VASP)~\cite{kresse_efficient_1996}.
The exchange-correlations were included using a generalized gradient approximation (GGA), and the spin-orbit coupling (SOC) was included self-consistently~\cite{kresse_ultrasoft_1999,perdew_generalized_1996}.
The on-site Coulomb interaction was added for Pr $f$-electrons within the GGA+U scheme with $U_{\textrm{eff}} = 6$~eV.
A Wannier tight-binding Hamiltonian was obtained from the ab-initio results, using the VASP2WANNIER90 interface, which was subsequently used in our topological properties calculations~\cite{marzari_maximally_1997}.

\emph{Second Harmonic Generation}-- The second-harmonic-generation (SHG) data were taken at normal incidence on the [101] face of as-grown crystals for incoming/outgoing wavelengths $1500/750$~nm as a function of the incoming field polarization and measured for emitted light polarized parallel to [010] crystalline axis.
In this geometry, all bulk contributions to SHG from a $I4_1/amd$ space group are forbidden, including bulk magnetic dipolar, electric quadrupolar, and electric-field induced SHG.

\section{\label{results} Results and Discussions}

\subsection{Crystal Structure}
\begin{figure}
\includegraphics[width=0.46\textwidth]{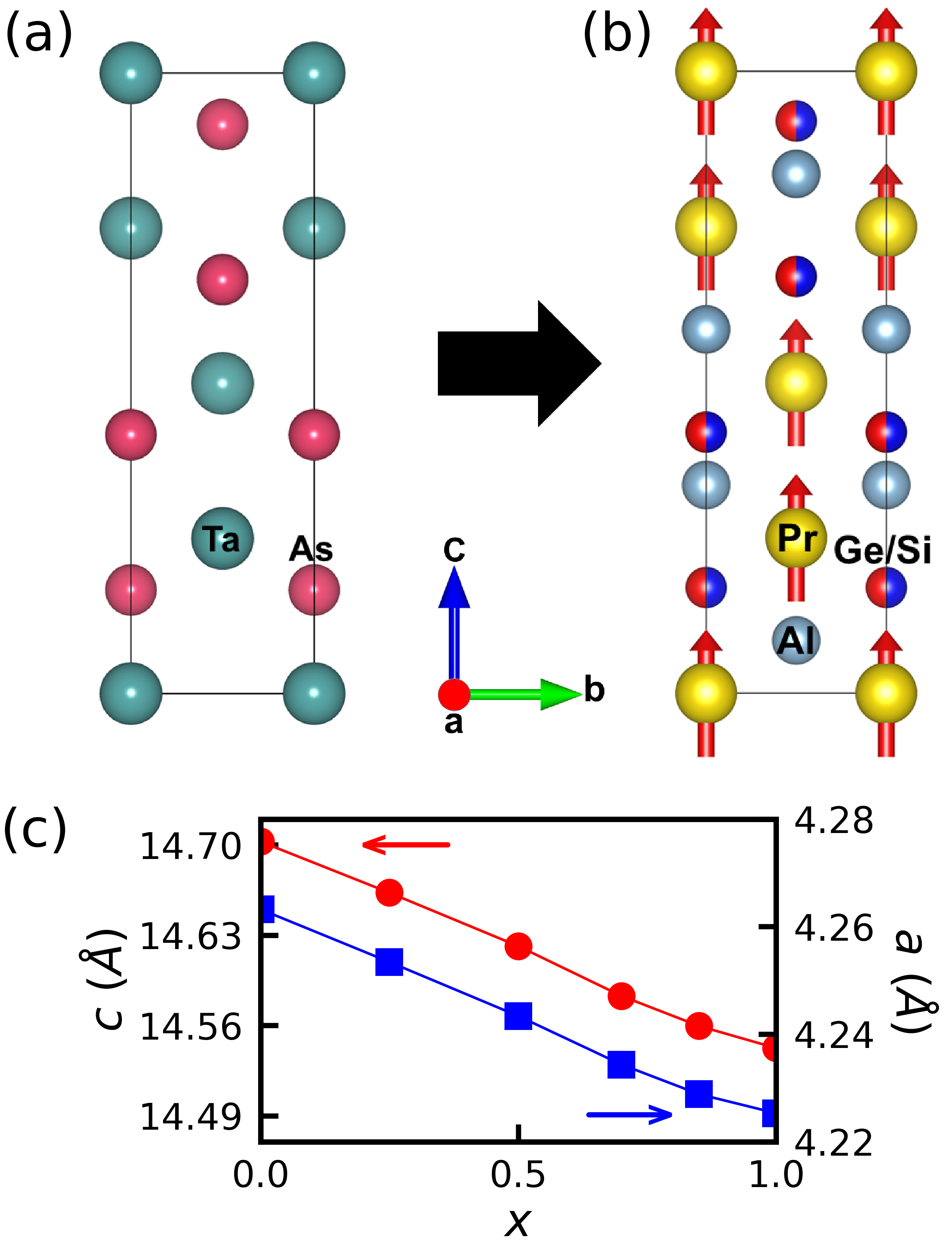}
\caption{\label{TP} (a) Crystal structure of TaAs in the space group I$4_1$md ($\#109$).
(b) The unit cell of PrAlGe$_{1-x}$Si$_x$, which is similar to TaAs but with additional Al atoms.
(c) Continuous change of lattice parameters as a function of $x$ among PrAlGe$_{1-x}$Si$_x$ compounds.
}
\end{figure}
PrAlGe and PrAlSi are both WSMs due to broken inversion symmetry~\cite{chang_magnetic_2018} similar to the archetypal WSM, TaAs~\cite{lv_experimental_2015,yang_weyl_2015}.
However, they undergo an FM transition at $T_C=15$-$17$~K unlike TaAs which remains non-magnetic at all temperatures.
As shown in Fig. \ref{TP}(a) and (b), TaAs and the PrAlGe$_{1-x}$Si$_x$ crystallize in the same noncentrosymmetric space group (I$4_1$md) with the important difference that the Pr atoms in PrAlGe$_{1-x}$Si$_x$ provide a net magnetic moment along the $c$-axis below $T_C$.
Furthermore, a solid solution of Ge and Si is realized in PrAlGe$_{1-x}$Si$_x$ which results in a continuous change of lattice parameters as seen in Fig.~\ref{TP}(c).
The lattice parameters in Fig.~\ref{TP}(c) are obtained from the Rietveld refinement of the powder X-ray diffraction data in the non-centrosymmetric space group I$4_1$md.
The point group C$_{4v}$ is confirmed by SHG refinements in Appendix \ref{app_SHG}.
This structure is characteristic of the entire $R$AlSi(Ge) family ($R=$ rare-earth), and generally leads to the appearance of Weyl nodes in their band structure~\cite{chang_magnetic_2018} as observed in LaAlGe,~\cite{xu_discovery_2017} CeAlGe,~\cite{hodovanets_single-crystal_2018} PrAlGe,~\cite{meng_large_2019} and CeAlSi$_{0.3}$Ge$_{0.7}$.~\cite{suzuki_singular_2019}.
As we will see, the number and positions of the Weyl nodes are similar in \PAGS\ at different $x$ but the Fermi surface significantly changes across the series, giving rise to two regimes of AHE in the PrAlGe$_{1-x}$Si$_x$ family.
\subsection{Magnetic Properties}
\begin{figure}
\includegraphics[width=0.46\textwidth]{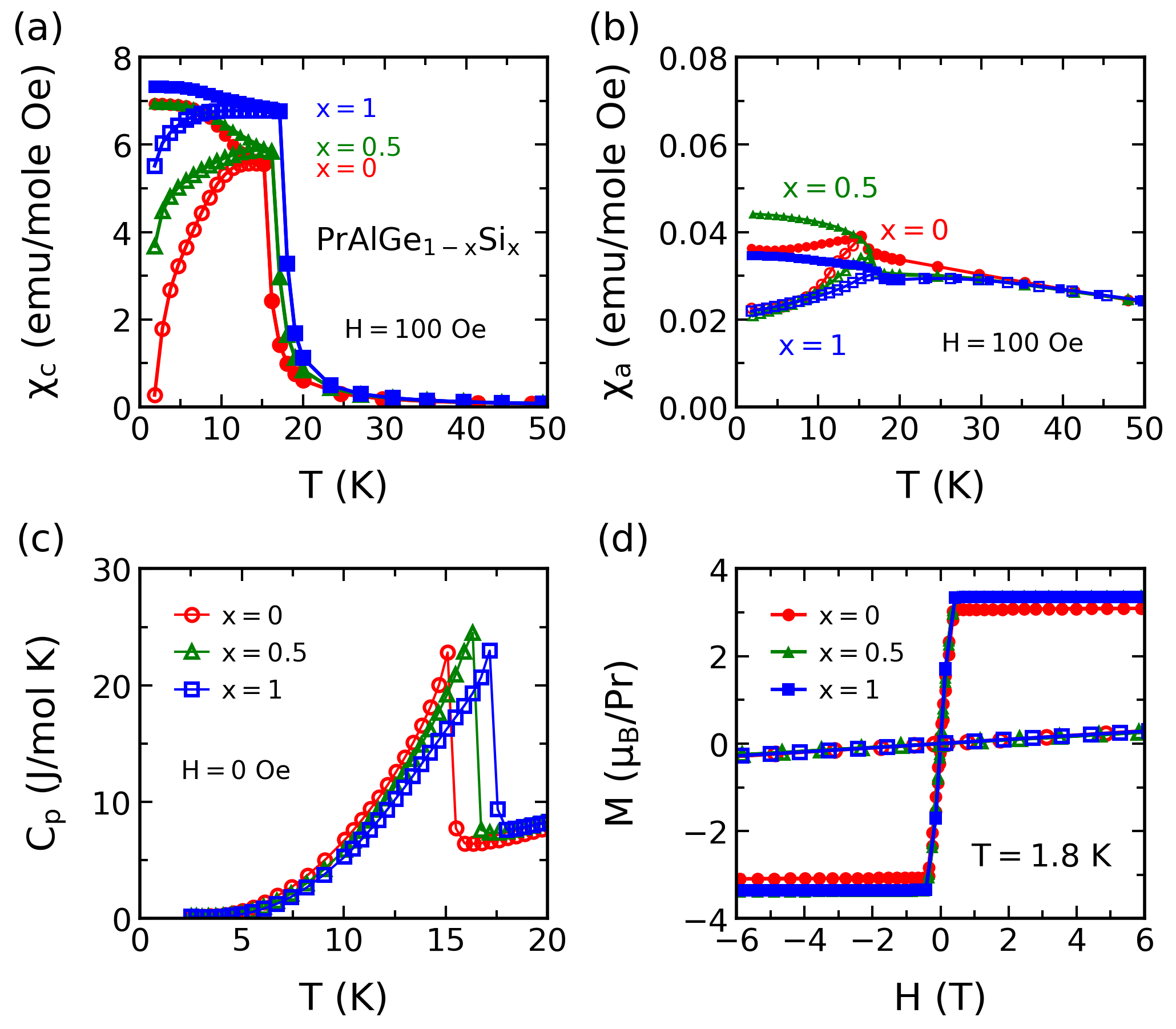}
\caption{\label{MH} (a) Magnetic susceptibility measured as a function of temperature with field parallel to the $c$-axis ($\chi_c$).
Red circles, green triangles, and blue squares represent the data for PrAlGe, PrAlGe$_{0.5}$Si$_{0.5}$, and PrAlSi, respectively (in all four panels).
Empty and full symbols correspond to zero-field-cooling (ZFC) and field-cooling (FC), respectively.
(b) Susceptibility data with field parallel to the $a$-axis ($\chi_a$).
Notice the $y$-scale is 100 times smaller than in panel (a) due to the Ising-like magnetic anisotropy.
(c) Heat capacity as a function of temperature. The peaks correspond to $T_C$.
(d) Magnetization as a function of field parallel to $c$- and $a$-axes (full and open symbols). Note that the coercive field is less than 0.1~T, so the hysteresis loop is not visible on this scale.
}
\end{figure}
A combination of magnetization and heat capacity measurements reveal the FM order, hence the breaking of time-reversal symmetry.
The magnetic properties are similar among \PAGS\ samples as seen in Fig.~\ref{MH} that shows representative data at $x=0,~0.5,$ and $1$.
The magnetic susceptibility is two orders of magnitude larger when measured with $H\|c$ ($\chi_c$ in Fig.~\ref{MH}(a)) compared to $H\|a$ ($\chi_a$ in Fig.~\ref{MH}(b)), indicating a strong Ising-like magnetic anisotropy.
Based on a Curie-Weiss analysis (see Appendix \ref{CW_ana}), the three samples have comparable Weiss temperatures $\Theta_W = 30-40$~K and effective moments $\mu_{\textrm{eff}} = 3.4-3.7$~$\mu_B$ as expected from Pr$^{3+}$ ($3.56~\mu_B$).
The FM transition temperature $T_C$ is evaluated from the peak in the heat capacity data which yields $T_C=15.1(2), 16.3(2),$ and $17.2(2)$~for PrAlGe, PrAlGe$_{0.5}$Si$_{0.5}$, and PrAlSi, respectively (see Fig.~\ref{MH}(c).)
The magnetization curves with $H\|c$ and $H\|a$ are compared in Fig.~\ref{MH}(d) where $M(H\|c)$ saturates at approximately $0.5$~T but $M(H\|a)$ does not.
This is consistent with $\chi_c \gg \chi_a$ in Fig.~\ref{MH}(a) and the Ising anisotropy depicted in Fig.~\ref{TP}(b) with the $c$-axis as the magnetic easy-axis.
The saturated moment for all \PAGS\ samples is approximately $3.3~\mu_B$/Pr.
In summary, the magnetic properties of \PAGS\ samples are nearly identical.

\subsection{Band Structure}
\begin{figure}
\includegraphics[width=0.46\textwidth]{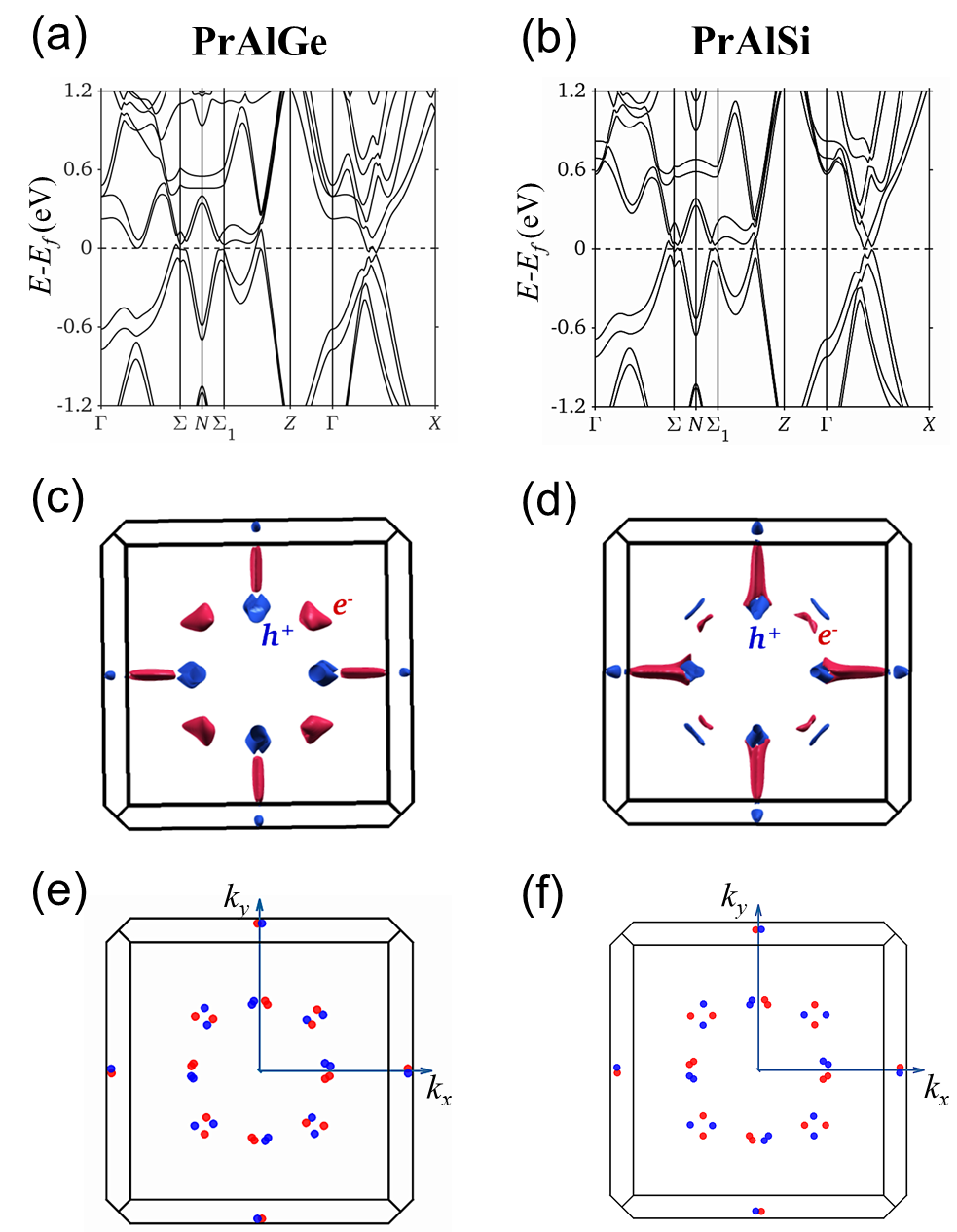}
\caption{\label{FS} A comparison is made between the band structures (a,b), Fermi surfaces (c,d), and Weyl nodes (e,f) in PrAlGe and PrAlSi, respectively. Both noncentrosymmetric space group symmetry and FM order are considered in the calculations shown here.
}
\end{figure}
We incorporated the crystalline and magnetic structures of PrAlGe and PrAlSi in our DFT calculation to arrive at realistic band structures.
We compare the calculated band structures, Fermi surfaces, and Weyl nodes in PrAlGe (Figs.~\ref{FS}(a,c,e)) and PrAlSi (Figs.~\ref{FS}(b,d,f)).
The band structure of both systems in Figs.~\ref{FS}(a,b) include tilted crossings near $\Sigma$ and $\Sigma_1$ characteristic of type-II Weyl semimetals.
The hole pocket between $\Sigma_1$ and $Z$ is visibly larger in PrAlGe than that in PrAlSi.
The Fermi surface is visualized for both PrAlGe and PrAlSi in Figs.~\ref{FS}(c,d) to highlight the larger size and the more isotropic shape of the hole pockets in PrAlGe compared to PrAlSi.
Although the Fermi surfaces are quite different between the two compounds, their Weyl node structures as shown in Figs.~\ref{FS}(e,f) are quite similar.
Both compounds have 40 Weyl nodes located at similar locations in the Brillouin zone.
Note that, the preservation of topological properties with the substitution of elements is not always guaranteed\cite{yang2017extreme}, and such a preservation in the PrAlGe$_{1-x}$Si$_x$ family makes it a great platform to study the competition between intrinsic and extrinsic mechanisms of AHE.

\subsection{Electronic Properties}
\begin{figure}
\includegraphics[width=0.46\textwidth]{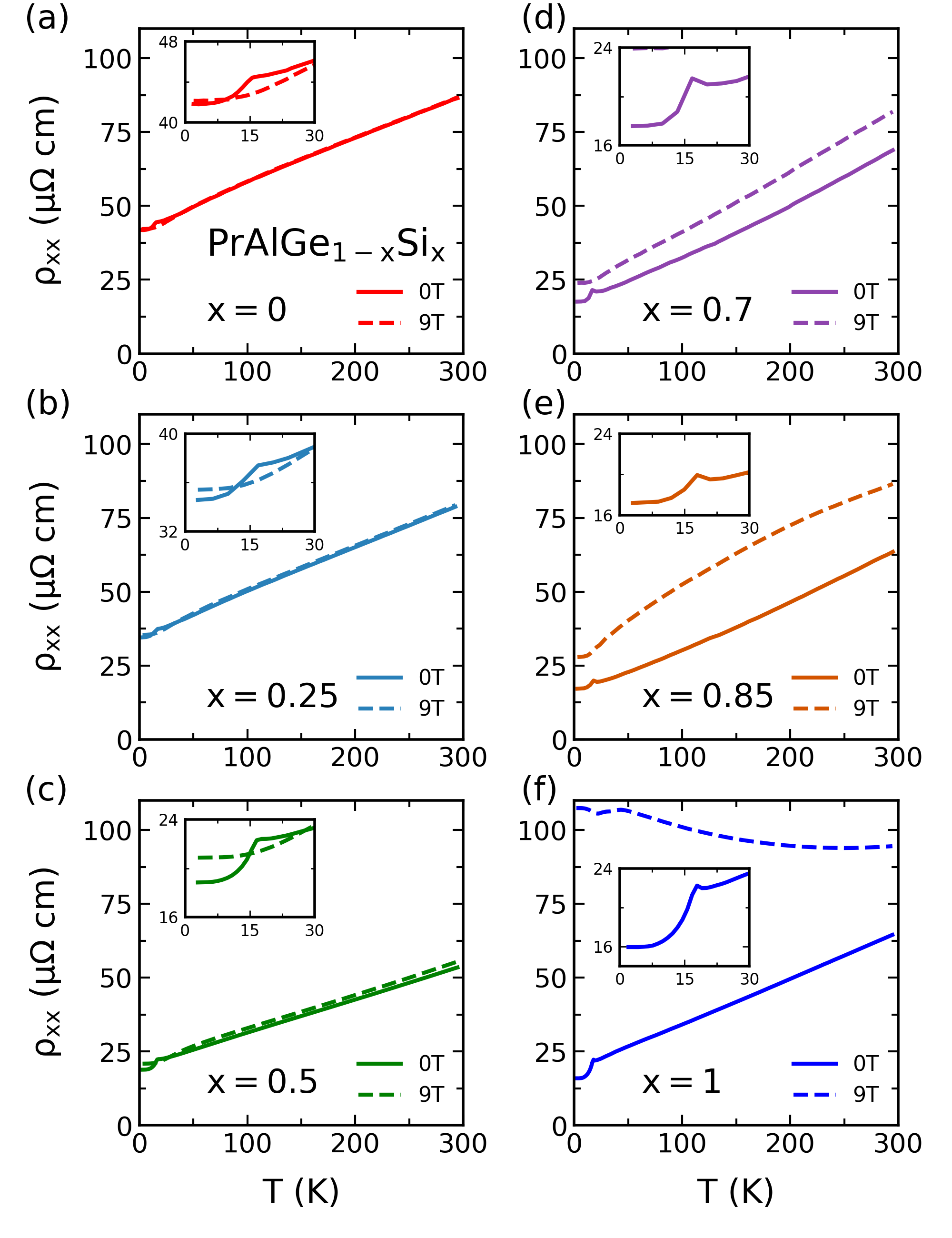}
\caption{\label{RT} Longitudinal resistivity $\rho_{xx}$ plotted as a function of $T$ in each \PAGS\ sample at both $H=0$ and $9$~T. The measurement was done with $I \| a$ and $H \| c$.
The Si content for each sample ($x$) is quoted in the corresponding panels (a to f).}
\end{figure}
Consistent with the change of the Fermi surface, we observe considerable changes in the magnetoresistance $\left(\textrm{MR}(\%)=100\times \frac{\rho_{xx}(H)-\rho_{xx}(0)}{\rho_{xx}(0)}\right)$ within the \PAGS\ series.
Figures~\ref{RT}(a) and ~\ref{RT}(f) show very different temperature dependences of MR between the end members, PrAlGe and PrAlSi.
The longitudinal resistivity $\rho_{xx}$ is measured for each sample at both $H=0$ (solid line) and $9$~T (dashed line).
PrAlGe shows a nearly field-independent $\rho_{xx}$ at $T>T_C$, thus a negligible MR.
A weak negative MR is observed near $T_c$ which is due to the magnetic fluctuations.
In contrast, PrAlSi shows a considerably larger $\rho_{xx}$ at $H=9$~T compared to the zero-field data, thus a strong MR at all temperatures from 1.8 to 300~K.
A continuous evolution of the MR is observed between these two limits in the rest of the \PAGS\ samples in Figs.~\ref{RT}(b-e).

\begin{figure}
\includegraphics[width=0.47\textwidth]{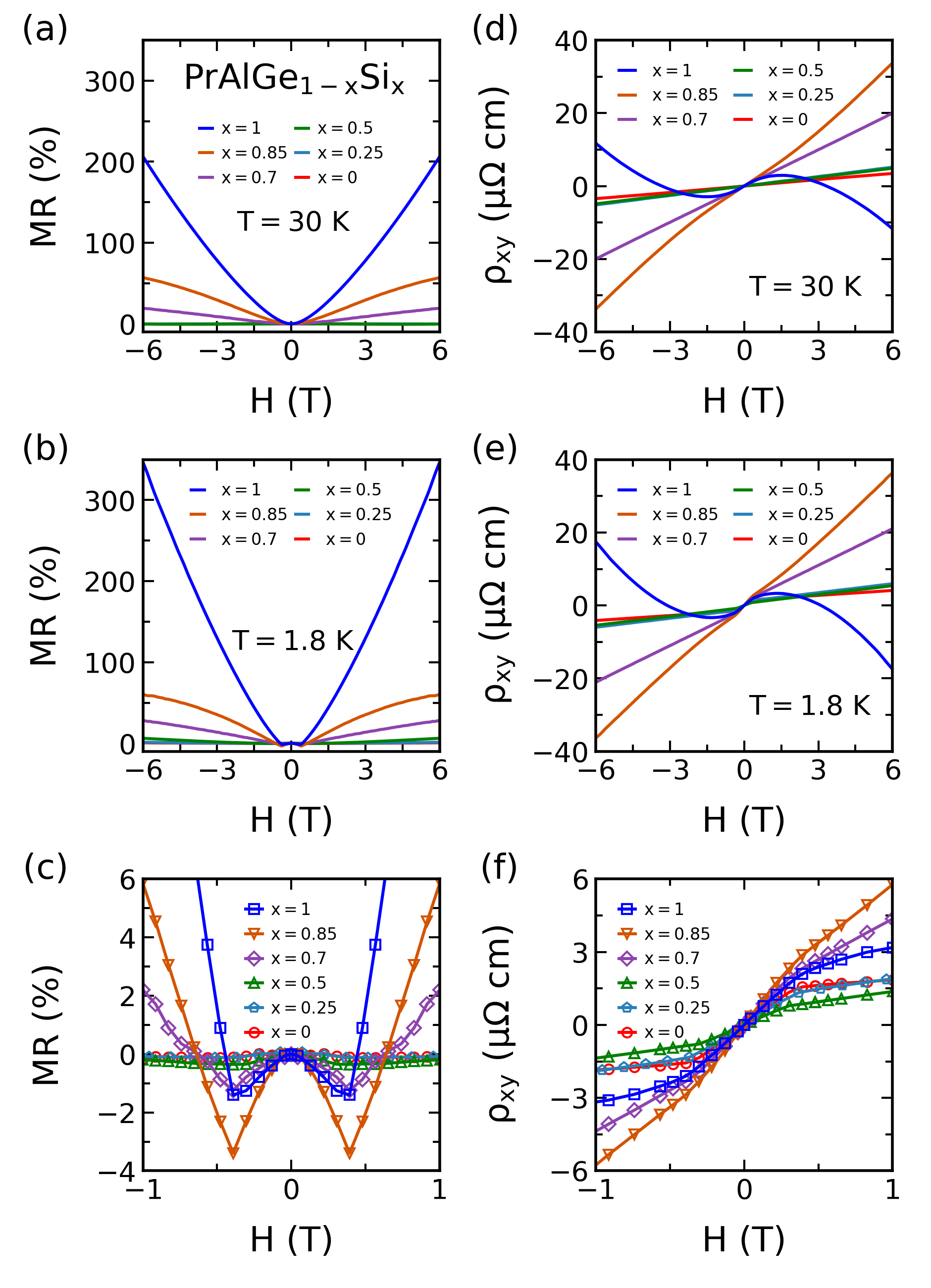}
\caption{\label{RH} (a) Transverse magnetoresistance in the \PAGS\ samples ($\textrm{MR}(\%)=100\times \left[\rho_{xx}(H)-\rho_{xx}(0)\right]/\rho_{xx}(0)$) measured as a function of field at $T=30$ K ($>T_C$). The current is along $a$-axis and the applied field is along $c$-axis ($z$) in all panels.
(b) MR at $T=1.8$ K ($<T_C$).
(c) A zoom-in view of panel (b) below 1~T.
(d) Hall resistivity ($\rho_{xy}$) measured as a function of field at $T=30$ K ($>T_C$).
(e) $\rho_{xy}$ at $T=1.8$ K ($<T_C$).
(f) A zoom-in view of panel (e) below 1~T.
}
\end{figure}
We present field dependences of both MR and Hall effect ($\rho_{xy}$) in Fig.~\ref{RH}.
A weak MR is observed in PrAlGe as a function of field at $T=30$~K (above $T_C$) in Fig.~\ref{RH}(a) cosistent with Fig.~\ref{RT}(a).
The MR gradually increases with increasing $x$ in the \PAGS\ series.
Eventually, the MR in PrAlSi ($x=1$) becomes 100 times larger than the MR in PrAlGe ($x=0$).
This behavior is more pronounced at $T=1.8$~K (below $T_C$) in Fig.~\ref{RH}(b).
We zoom in the low-field MR data in Fig.~\ref{RH}(c) to show the negative MR due to magnetic fluctuations in all samples.
Although the negative MR is more pronounced in samples with higher $x$ at $T=1.8$ K, it never exceeds $4\%$ and disappears at $T>2T_C$ as seen in Fig.~\ref{RT}.
Figure~\ref{RH}(d) shows a variation of the Hall resistivity $\rho_{xy}(H)$ between different samples at $T=30$~K (above $T_C$).
The slope of the Hall resistivity $d\rho_{xy}/dH$ is small and positive at $H>1$~T in PrAlGe; it gradually increases with increasing $x$ and becomes significantly larger in PrAlGe$_{0.15}$Si$_{0.85}$.
Eventually, it shows an abrupt downturn in PrAlSi (at $H>1$~T).
This behavior is consistent with our DFT calculations that show smaller hole pockets with increasing Si-content $x$ in Fig.~\ref{FS}.
A gradual weakening of the antibonding orbital overlaps between the $p$-orbitals of Al and Ge/Si with increasing $x$ in PrAlGe$_{1-x}$Si$_{x}$ leads to smaller hole and larger electron Fermi surfaces.
The $\rho_{xy}$ behavior at high fields remains unchanged at $T=1.8$~K (below $T_C$) as seen in Fig.~\ref{RH}(e).
A zoom-in view at low-fields in Fig.~\ref{RH}(f) reveals the anomalous Hall effect (AHE) in all samples at $T=1.8$~K characterized by a rapid increase of $\rho_{xy}(H)$ until $H=0.5$~T followed by a linear field dependence from $H=0.5$ to $1$~T.
In the next section, we examine the AHE in detail and reveal a transition from intrinsic to extrinsic AHE in \PAGS.

\subsection{Anomalous Hall Effect}
\begin{figure}
\includegraphics[width=0.46\textwidth]{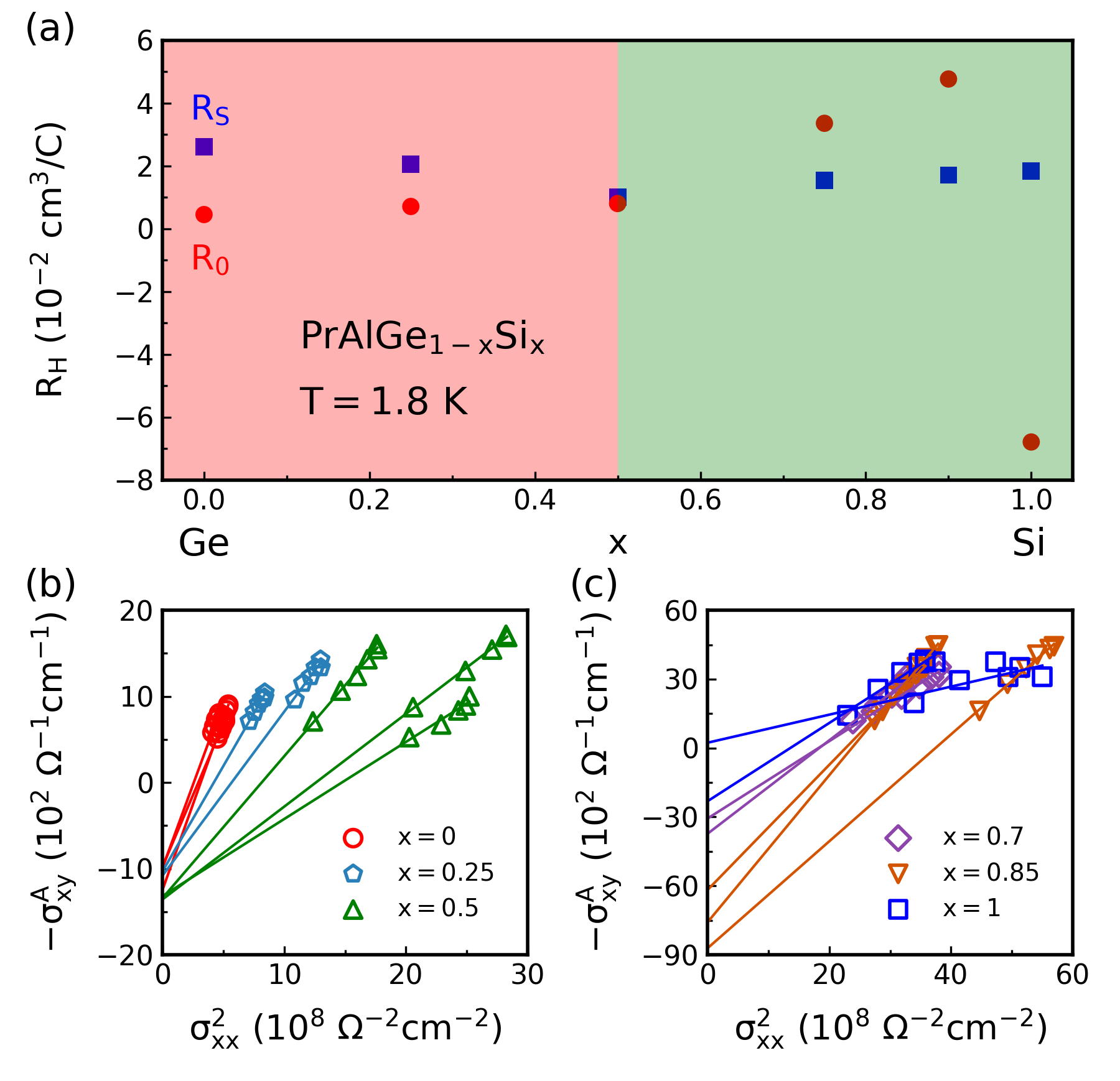}
\caption{\label{AHE} (a) Ordinary and anomalous Hall coefficients are plotted as red circles ($R_0$) and blue squares ($R_S$), respectively, as a function of $x$ in \PAGS\ at $T=1.8$~K.
The red ($x \leq 0.5$) and green ($x > 0.5$) backgrounds distinguish two regions where the scaling analysis yields the same intercept (b) or different intercepts (c) according to Eq.~\ref{sigmaxx}.
}
\end{figure}
We present two separate analyses to investigate the AHE.
The first analysis is based on Eq.~\ref{hall} to differentiate the relative magnitudes of the ordinary and anomalous Hall coefficients ($R_0$ and $R_S$) in \PAGS.
The details of extracting $R_0$ and $R_S$ are presented in Appendix~\ref{app_1st_ana}.
We plot both $R_0$ and $R_S$ as a function of $x$ in Fig.~\ref{AHE}(a) to reveal a crossing between the magnitudes of $R_0$ and $R_S$ at $x=0.5$ so that $|R_0|/|R_S|<1$ at $x\le 0.5$ but $|R_0|/|R_S|>1$ at $x> 0.5$.
The magnitude of $R_S$ moderately decreases with increasing $x$ at $x\leq 0.5$ and remains nearly unchanged afterwards.
Whereas $R_S$ shows only mild variations, $R_0$ shows considerable variations with $x$ due to the change of Fermi surface shown in Fig.~\ref{FS}.
$R_0$ is positive and increases slowly between $x=0$ and $0.5$, then increases rapidly until $x=0.85$, and finally becomes negative abruptly at $x=1$.
Both the different $|R_0|/|R_S|$ ratios and different behaviors of the two Hall coefficients at $x\le0.5$ and $0.5$ suggest a transition from one regime to another at $x=0.5$.
These observations motivate our second analysis.
We follow the empirical analysis which was first proposed by Tian \emph{et al.}~\cite{tian2009proper} on iron thin films and was later justified theoretically~\cite{shitade2012anomalous,hou2015multivariable}.
The analysis assumes a material not in the clean limit where the residual resistivity $\rho_{xx0}$ plays an important role and the phonon scattering does not.
These conditions are satisfied in \PAGS\ where RRR~$<4$ and the AHE occurs below $17$~K so phonon scattering is negligible.
Under such circumstance, the AHE can be described as
\begin{equation}
\label{rhoxx}
\rho_{xy}^A=\left(\alpha\rho_{xx0}+\beta\rho_{xx0}^2\right)+b\rho_{xx}^2
\end{equation}
where the coefficients $\alpha$, $\beta$, and $b$ parametrize the skew-scattering, side-jump, and intrinsic contributions to $\rho_{xy}^A$.
The first two parameters depend on the impurity scattering in a specific sample, but the parameter $b$ is independent of scattering.
In general, the side-jump mechanism could also contribute to $b$ through the same $\rho_{xx}^2$ dependence\cite{hou2015multivariable,weischenberg2011ab}.
However, in a material with topological band structures, we expect the intrinsic contribution to be dominant.
Dividing both sides of Eq.~\ref{rhoxx} by $\rho_{xx}^2$ (and assuming that $\rho_{xx} \gg \rho_{xy}^A$)\footnote{In the limit $\rho_{xx} \gg \rho_{xy}^A$, the relations $\sigma_{xy}^A \sim \rho_{xy}^A/\rho_{xx}^2$ and $\sigma_{xx} \sim 1/\rho_{xx}^2$ are valid. In PrAlGe$_{1-x}$Si$_x$ this is the case as can be seen in the resistivity data ($\rho_{xy}^A$ in Fig. \ref{RH}(f) and $\rho_{xx}$ in Fig. \ref{RT}.)}
yields:
\begin{equation}
\label{sigmaxx}
\sigma_{xy}^A=-\left(\alpha\sigma_{xx0}^{-1}+\beta\sigma_{xx0}^{-2}\right)\sigma_{xx}^2-b
\end{equation}
where $\sigma_{xx0}=1/\rho_{xx0}$ is the residual conductivity and $\sigma_{xy}^A=-\rho_{xy}^A/\rho_{xx}^2$ is the anomalous Hall conductivity (AHC).
The first term in Eq.~\ref{sigmaxx} depends on the residual conductivity and is sample dependent.
However, the second term ($b$) is sample independent and constitutes the intrinsic contribution to the AHC.
Following Eq.~\ref{sigmaxx}, we measured two to three samples for each composition of \PAGS, and determined $b$ from the intercept of a linear fit to $\sigma_{xy}^A$ as a function of $\sigma_{xx}^2$.
For example, the three data sets with green triangles in Fig.~\ref{AHE}(b) correspond to three samples of PrAlGe$_{0.5}$Si$_{0.5}$.
Their respective linear fits have  different slopes showing different disorder levels, thus different $\alpha$ and $\beta$ fitting parameters in Eq.~\ref{sigmaxx}.
However, all three lines end at the same intercept ($b$) in the limit of $\sigma_{xx}\to 0$.
The convergence of all linear fits strongly suggests an intrinsic mechanism for the AHE, which does not depend on the details of disorder level and only cares about the overall band structure.
Interestingly, $b$ seems to be the same in the three compositions $x=0,~0.25$, and $0.5$ which is reasonably justified by the similar nodal structure of all \PAGS\ as illustrated in Fig.~\ref{FS}(c,f).
Also, the magnitude of $\sigma_{xy}^{int} = -b \approx 10^3 \, \Omega^{-1} \text{cm}^{-1}$ is consistent with the magnitude of the resonant AHE caused by the intrinsic mechanism~\cite{onoda2006intrinsic,miyasato2007crossover}.
Therefore, Fig.~\ref{AHE}(b) suggests a universal intrinsic AHE in samples with $x\le0.5$.
In contrast to Fig.~\ref{AHE}(b), Fig.~\ref{AHE}(c) shows that the parameter $b$ varies randomly among samples with $x>0.5$, hence the absence of a universal $\sigma_{xy}^{int}$.
The failure of the scaling analysis suggests a predominantly extrinsic contribution to the AHC.
Thus, we conclude that the AHE evolves from an intrinsic regime ($x \leq 0.5$) to an extrinsic one ($x>0.5$) in PrAlGe$_{1-x}$Si$_x$, despite similar Weyl node structures in both end members, PrAlGe and PrAlSi.

\subsection{Discussion}
Finally, we compare the AHE in PrAlGe$_{1-x}$Si$_x$ family of FM WSMs to the one in Co$_3$Sn$_2$S$_2$, which is also an FM WSM. Co$_3$Sn$_2$S$_2$ exhibits a very large AHE but its Weyl nodes are located at 60 meV above $E_F$\cite{liu_giant_2018}.
In both PrAlGe and PrAlSi, we found that there are 40 Weyl nodes in their band structures (see Appendix \ref{WPs} for details).
Besides, most of them are less than 60~meV away from $E_F$; some of them are even only few meV away in PrAlSi.
Although the Weyl nodes in PrAlGe$_{1-x}$Si$_x$ are closer to $E_F$, compared to those in Co$_3$Sn$_2$S$_2$, PrAlGe$_{1-x}$Si$_x$ shows smaller AHC.
This comparison does not follow the common wisdom that since Weyl nodes are like singularities of Berry curvatures, if they are located near $E_F$ the system should exhibit a huge AHC\cite{burkov_anomalous_2014}.
Also, the extrinsic contribution seems to be significant for all compositions among PrAlGe$_{1-x}$Si$_x$ even in the intrinsic regime with well-defined $\sigma_{xy}^{int}$. In Fig. \ref{AHE}(b), it seems that the extrinsic contribution actually competes against $\sigma_{xy}^{int}$, and eventually changes the sign of $\sigma_{xy}$ within the observable range.
Our finding thus reveals the potential inertness of Weyl nodes' contributions to AHE even if the Weyl nodes are very close to $E_F$. More studies on different FM WSMs will help clarify this point and the hierarchy of different contributions to AHE in systems hosting topological band structures.

\pagebreak

\begin{acknowledgments}
We thank Chunli Huang and Hiroaki Ishizuka for helpful discussions.
The work at Boston College was funded by the National Science Foundation through NSF/DMR-1708929.
The work at Northeastern University was supported by the US Department of Energy (DOE), Office of Basic Energy Sciences, Grant No. DE-FG02-07ER46352, and benefited from Northeastern University's Advanced Scientific Computation Center and the National Energy Research Scientific Computing Center through DOE Grant No. DE-AC02-05CH11231.
The National High Magnetic Field Laboratory is supported by the National Science Foundation through NSF/DMR-1644779 and the State of Florida.
\end{acknowledgments}

\appendix


\section{EDX Results}\label{app_EDX}
\begin{table*}[b]
\caption{\label{T1} EDX results for different PrAlGe$_{1-x}$Si$_x$ samples. The atomic weights derived from EDX spectra are normalized to the Pr content. These results are subject to an error of $\pm0.05$.}
\begin{tabularx}{\textwidth}{X|XXXXX}
$y$               & Pr & Al & Ge & Si & Si/(Si+Ge) ($x$)  \\
\hline
\hline
0   & 1.00 & 1.00 & 0.00 & 1.07 & 1.00  \\
0.5   & 1.00 & 0.90 & 0.47 &0.49 & 0.51  \\
0.75   & 1.00 & 1.08 & 0.37 & 0.82 & 0.69  \\
0.9   & 1.00 & 0.95 & 0.15 & 0.87 & 0.85  \\
1   & 1.00 & 0.82 & 0.00 & 0.86 & 1.00  \\
\end{tabularx}
\end{table*}
We performed an EDX analysis to evaluate the Si-content ($x$) in \PAGS\ accurately.
The first column in Table~\ref{T1} shows the mole ratio used in the crystal growth (Pr:Al:Ge:Si = 1:10:$y$:$1-y$).
The last column shows the final Si-content ($x$) in each \PAGS\ crystal.
For $y=0$, 0.5 and 1, $x=y$.
For $y=0.75$ and 0.9, $x=0.7$ and 0.85, respectively.

\section{Second Harmonic Generation}\label{app_SHG}
\begin{figure}
\includegraphics[width=0.46\textwidth]{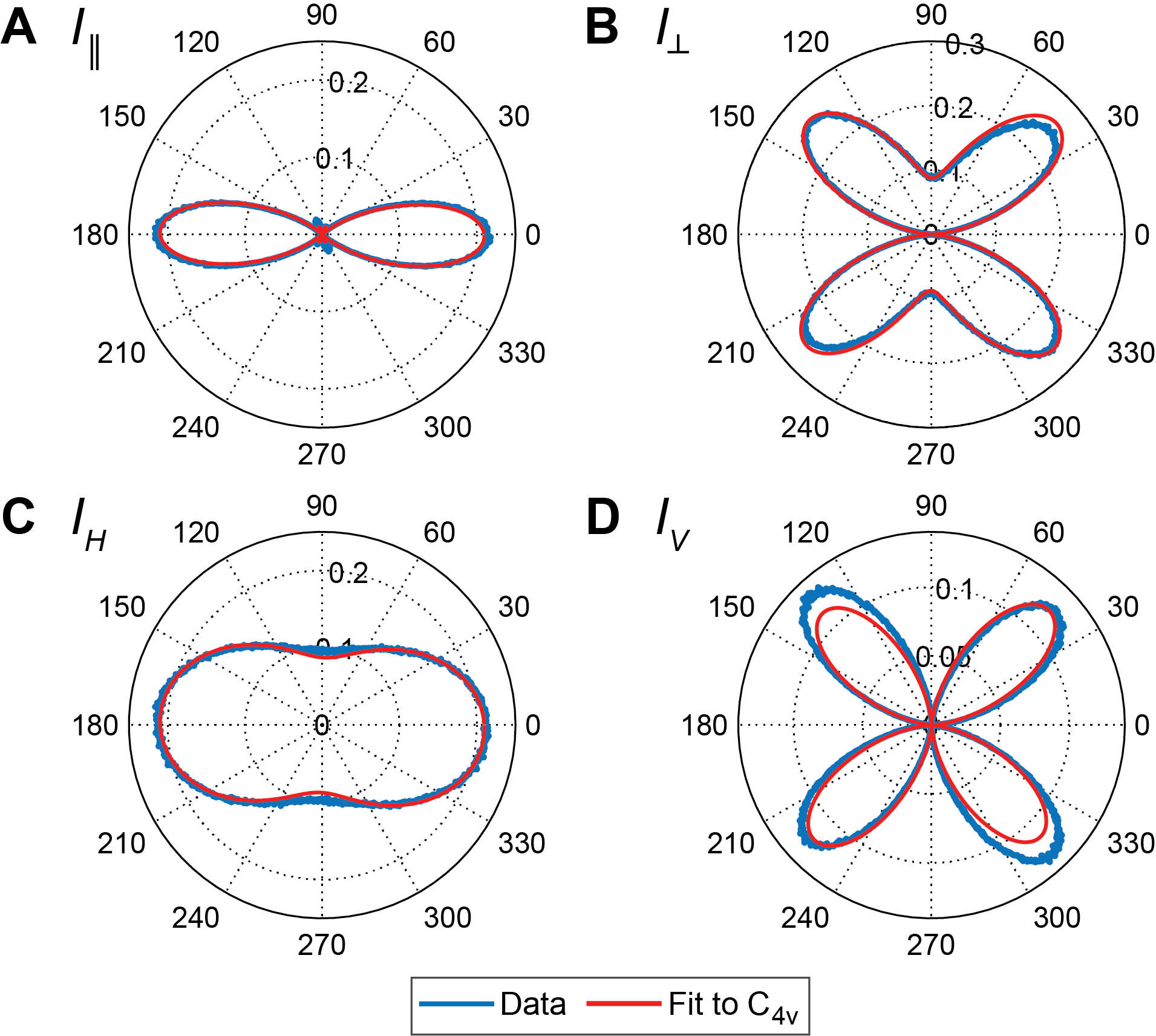}
\caption{\label{SHG} The SHG data for incoming wavelength 1500~nm, outgoing wavelength 750~nm, and fits to bulk electric dipolar SHG in the $C_{4\textrm{v}}$ point group as given by Eqs.~[\ref{eq:eee1}-\ref{eq:eee4}] for (\textbf{A}) $I_\parallel$, (\textbf{B}) $I\perp$, (\textbf{C}) $I_V$, and (\textbf{D}) $I_H$.
}
\end{figure}
The SHG data were fit to functions appropriate to four different experimental configurations:
incoming polarization rotating, output polarizer fixed with polarization parallel to the crystalline $[010]$ axis, denoted $I_{H}(\phi)$;
incoming polarization rotating, output polarizer fixed with polarization parallel to the $[10\overline{1}]$ axis, denoted $I_{V}(\phi)$;
incoming polarization rotating, outgoing polarizer rotated at 0$^\circ$ angle relative to the incoming polarization, denoted $I_{\parallel}(\phi)$;
and incoming polarization rotating, outgoing polarizer rotated with polarization axis at 90$^\circ$ angle relative to the incoming polarization, denoted $I_{\perp}(\phi)$.
In the electric dipole approximation, the mathematical forms of these various responses for the $[101]$ crystal face in the $I4_1md$ space group ($C_4v$ point group) are given by
\begin{widetext}
\begin{align}
&I^{eee}_{\parallel}(\phi) = \frac{1}{32} \cos ^2(\phi ) \left[ (-2 \chi^{eee}_{xxz}-\chi^{eee}_{zxx}+\chi^{eee}_{zzz})\cos (2 \phi )+6 \chi^{eee}_{xxz}+3 \chi^{eee}_{zxx}+\chi^{eee}_{zzz}\right]^2\label{eq:eee1}\\
&I^{eee}_{\perp}(\phi) = \frac{1}{8} \sin ^2(\phi )\left[ (-2 \chi^{eee}_{xxz}+\chi^{eee}_{zxx}+\chi^{eee}_{zzz}) \cos ^2(\phi )+2 \chi^{eee}_{zxx} \sin ^2(\phi )\right]^2\label{eq:eee2}\\
&I^{eee}_{H}(\phi) = \frac{1}{8} \left[(2 \chi^{eee}_{xxz}+\chi^{eee}_{zxx}+\chi^{eee}_{zzz})\cos ^2(\phi )+2 \chi^{eee}_{zxx} \sin ^2(\phi )\right]^2\label{eq:eee3}\\
&I^{eee}_{V}(\phi) = 2\left[\chi^{eee}_{xxz} \sin(\phi ) \cos(\phi )\right]^2\label{eq:eee4}
\end{align}
\end{widetext}
The data were fit to expressions~[\ref{eq:eee1}-\ref{eq:eee4}] accounting for a rotation of the sample axes relative to the laboratory x-axis to produce excellent fits to the data, as seen in Fig.~\ref{SHG}.
The competing space group assignment $I4_1/amd$ (point group $D_{4h}$) is centrosymmetric and thus should not produce as strong of a SHG response as we have shown here.

\section{Curie-Weiss Analysis}\label{CW_ana}
\begin{figure}
\includegraphics[width=0.46\textwidth]{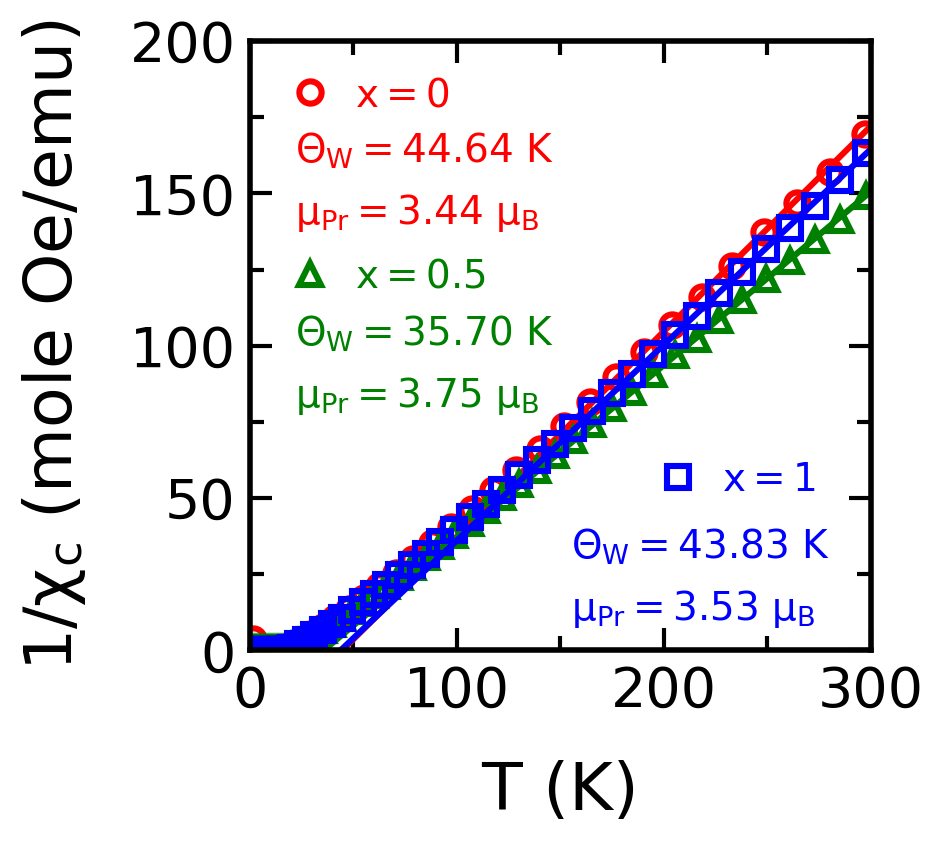}
\caption{\label{CW} Curie-Weiss analysis for PrAlGe$_{1-x}$Si$_x$ with $x=0$, 0.5, and 1.}
\end{figure}
The Curie-Weiss analysis was performed on PrAlGe$_{1-x}$Si$_x$ with $x=0$, 0.5 and 1.
The Curie-Weiss fit was made to the high temperature data ($T>150$ K) to extract the Weiss temperature $\Theta _W$ and the effective moment $\mu_{\text{Pr}}$ as seen in Fig. \ref{CW}.
Based on this analysis, different compositions in PrAlGe$_{1-x}$Si$_x$ family have similar $\Theta_W$ and $\mu_{\text{Pr}}$.

\section{The Analysis of $R_0$ and $R_S$}\label{app_1st_ana}

\begin{figure*}
\includegraphics[width=0.7\textwidth]{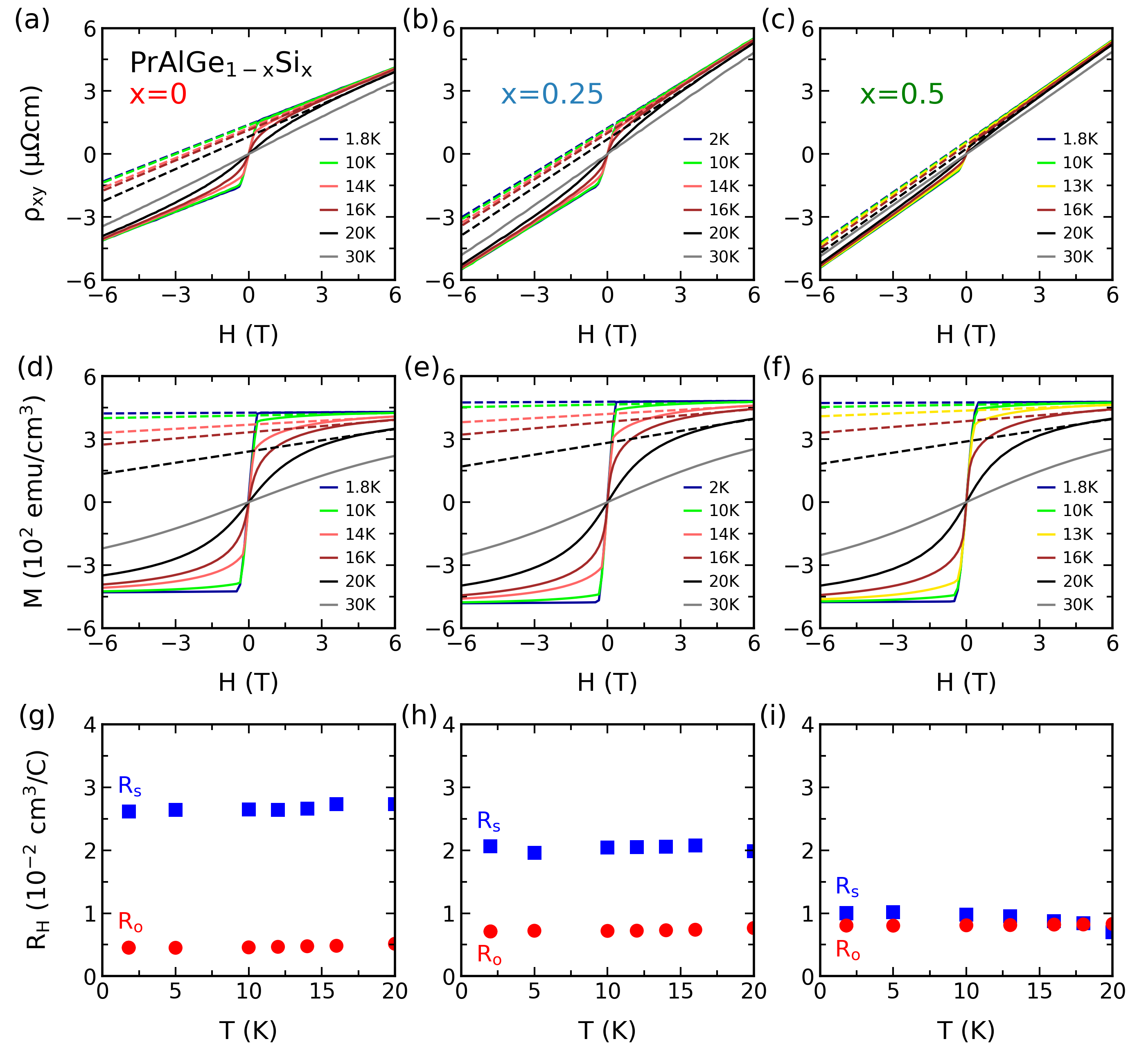}
\caption{\label{R0RS_1} Data used to extract $R_0$ and $R_S$ in PrAlGe$_{1-x}$Si$_x$ compounds for $x\leq 0.5$. (a-c) $\rho_{xy}$ measured at different temperatures. The y-intercepts of the fitting lines are extracted as $\rho_{xy}^A$. (d-f) Magnetization measured at different temperatures. Fitting lines are made to high-field part of the data. (g-i) $R_0$ and $R_S$ values at different temperatures. }
\end{figure*}

\begin{figure*}
\includegraphics[width=0.7\textwidth]{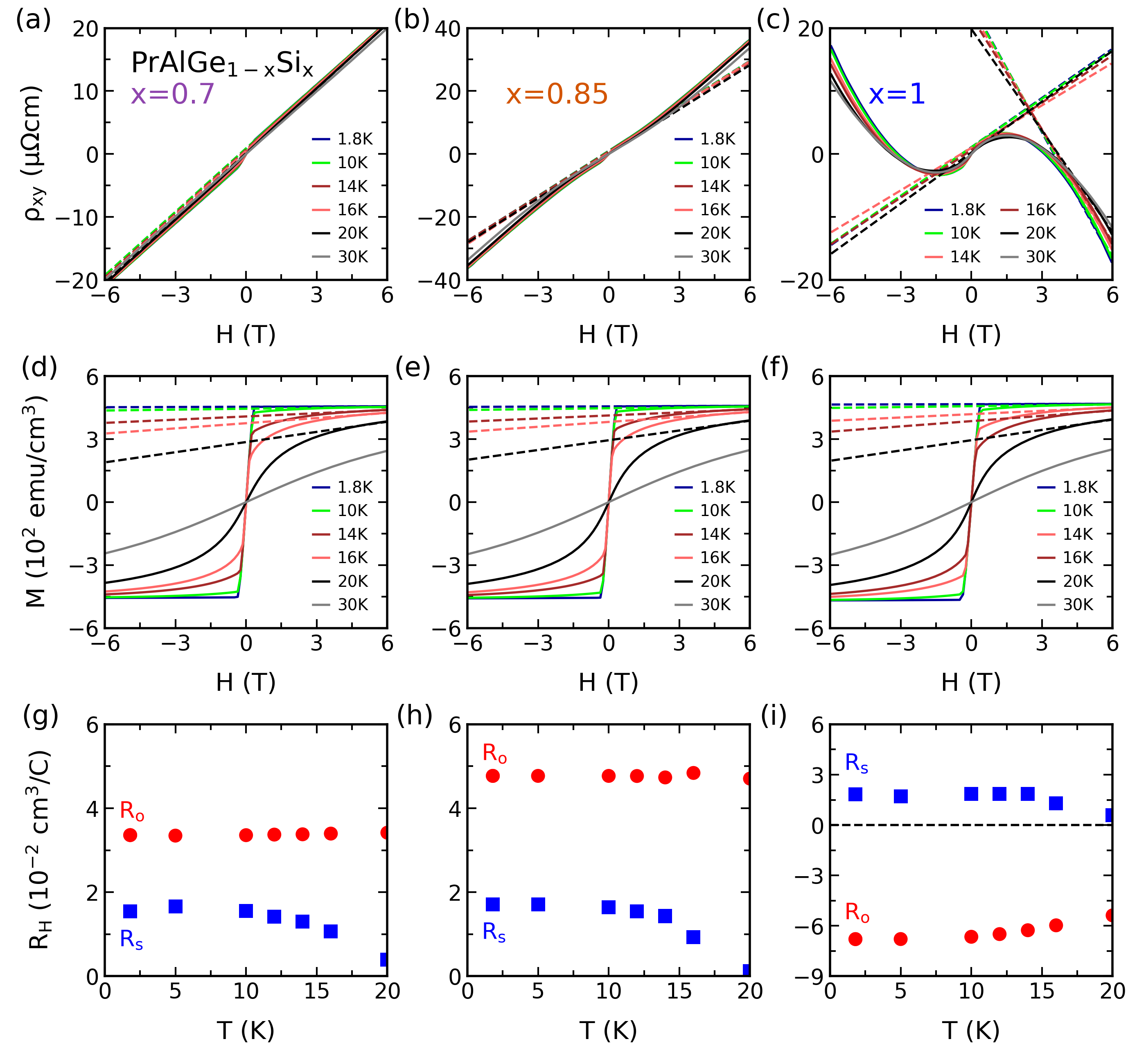}
\caption{\label{R0RS_2} Data used to extract $R_0$ and $R_S$ in PrAlGe$_{1-x}$Si$_x$ compounds for $x>0.5$. (a-c) $\rho_{xy}$ measured at different temperatures. The y-intercepts of the fitting lines are extracted as $\rho_{xy}^A$. (d-f) Magnetization measured at different temperatures. Fitting lines are made to high-field part of the data. (g-i) $R_0$ and $R_S$ values at different temperatures. }
\end{figure*}
Here, we show all the data required to extract $R_0$ and $R_S$ according to Eq.~\ref{hall} in Fig.~\ref{R0RS_1} and \ref{R0RS_2}.
Note that, for $x=0.85$ and $x=1$, $\rho_{xy}$ is not linear at high fields.
Thus, to extract $\rho_{xy}^A$ for these two compositions, the data between $H=0.5$ and $1$ T were fitted to a line, the intercept of which at $H=0$ T was extracted as $\rho_{xy}^A$.
For other compositions, the data at $H>4$ T were used for the fitting lines to extract $\rho_{xy}^A$.
The slopes of these lines were reported as $R_0$ except for $x=1$, where a second fitting line was made to high-field data to capture the behavior of the ordinary Hall effect.

\section{Quantum Oscillation and Calculated Anomalous Hall Conductivity}\label{app:QO_AHC}
\begin{figure}
\includegraphics[width=0.46\textwidth]{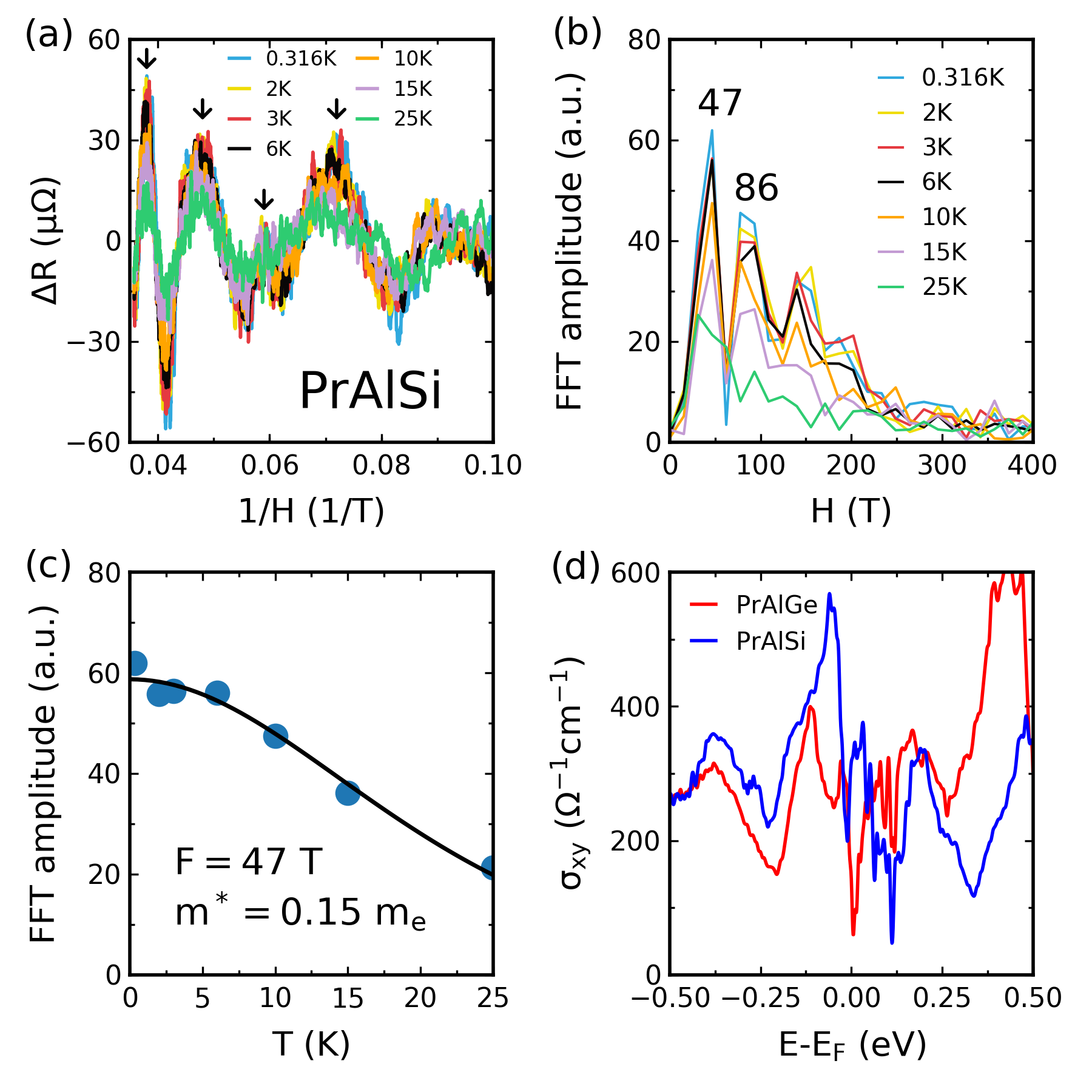}
\caption{\label{QO} (a) Shubnikov-de Haas oscillations in PrAlSi.
The field was applied along $c$-axis.
The arrows mark the peaks of the oscillations.
(b) The fast Fourier transform (FFT) of the data shown in (a).
(c) The effective mass extracted by the Lifshitz-Kosevich formula~\cite{shoenberg_magnetic_2009} for the Fermi surface corresponds to $F=47$~T.
(d) Calculated intrinsic AHC for both PrAlGe and PrAlSi.
}
\end{figure}
The results of quantum oscillation (QO) experiment and the calculation of AHC can be seen in Fig.~\ref{QO}.
Note that to match the frequency extracted from QO experiment with the one given by band structure calculation, the $E_F$ in the calculation has to be shifted up by 6~meV for PrAlSi.
At 6~meV in Fig.~\ref{QO}(d), the calculated AHC is about 300-400 $\Omega^{-1}$cm$^{-1}$ for PrAlSi.
Since the calculation here only considers the contribution from Berry curvature, there could be a discrepancy as shown in the literature~\cite{weischenberg2011ab}. 
Overall, the AHC calculation has the same magnitude as the one extracted from the experiment.

\section{Energies and Positions of the Weyl nodes in PrAlGe and PrAlSi}\label{WPs}

We looked into the band structures given by DFT calculations and searched for Weyl nodes, which are sources and drains of Berry curvatures. In both PrAlGe and PrAlSi, we identified 40 Weyl nodes; their positions in k-space and energies with respect to $E_F$ are listed thoroughly in Table. \ref{tab:WP_Ge} and \ref{tab:WP_Si}.
\begin{table*}[!htb]
    \caption{Energies and Positions of Weyl nodes in PrAlGe.}
    \label{tab:WP_Ge}
    \begin{minipage}{.5\linewidth}

      \centering
        \begin{tabularx}{0.97\textwidth}{p{0.6cm}|XXXXp{0.9cm}}
WP              & $k_x \ (\angstrom ^{-1})$ & $k_y \ (\angstrom ^{-1})$ & $k_z \ (\angstrom ^{-1})$ & $E$-$E_F$ (eV) & Charge  \\
\hline
\hline
1   & 0.34298 & 0.28801 & 0.00126 & -0.036 & -1  \\
2   & -0.28801 & 0.34298 & 0.00126 & -0.036 & -1  \\
3   & 0.28801 & -0.34298 & 0.00126 & -0.036 & -1  \\
4   & -0.34298 & -0.28801 & 0.00126 & -0.036 & -1  \\
5   & 0.04110 & -0.34697 & 0.28112 & 0.027 & -1  \\
6   & 0.34697 & 0.04110 & 0.28112 & 0.027 & -1  \\
7   & -0.34697 & -0.04110 & 0.28112 & 0.027 & -1  \\
8   & -0.04110 & +0.34697 & 0.28112 & 0.027 & -1  \\
9   & -0.24497 & -0.27041 & -0.00056 & -0.031 & -1  \\
10   & 0.27041 & -0.24497 & -0.00056 & -0.031 & -1  \\
11   & -0.27041 & 0.24497 & -0.00056 & -0.031 & -1  \\
12   & 0.24497 & 0.27041 & -0.00056 & -0.031 & -1  \\
13   & 0.77159 & 0.01209 & -0.01292 & 0.05 & +1  \\
14   & -0.01209 & 0.77159 & -0.01292 & 0.05 & +1  \\
15   & 0.01209 & -0.77159 & -0.01292 & 0.05 & +1  \\
16   & -0.77159 & -0.01209 & -0.01292 & 0.05 & +1  \\
17   & -0.02871 & 0.36769 & -0.29316 & 0.67 & -1  \\
18   & -0.36769 & -0.02871 & -0.29316 & 0.67 & -1  \\
19   & 0.36769 & 0.02871 & -0.29316 & 0.67 & -1  \\
20   & 0.02871 & -0.36769 & -0.29316 & 0.67 & -1  \\
\end{tabularx}
    \end{minipage}%
    \begin{minipage}{.5\linewidth}
      \centering

       \begin{tabularx}{0.97\textwidth}{p{0.6cm}|XXXXp{0.9cm}}
WP              & $k_x \ (\angstrom ^{-1})$ & $k_y \ (\angstrom ^{-1})$ & $k_z \ (\angstrom ^{-1})$ & $E$-$E_F$ (eV) & Charge  \\
\hline
\hline
21   & -0.34242 & 0.28818 & -0.00135 & -0.034 & +1  \\
22   & -0.28818 & -0.34242 & -0.00135 & -0.034 & +1  \\
23   & 0.28818 & 0.34242 & -0.00135 & -0.034 & +1  \\
24   & 0.34242 & -0.28818 & -0.00135 & -0.034 & +1  \\
25   & -0.02771 & -0.36309 & 0.29722 & 0.071 & +1  \\
26   & 0.36309 & -0.02771 & 0.29722 & 0.071 & +1  \\
27   & -0.36309 & 0.02771 & 0.29722 & 0.071 & +1  \\
28   & 0.02771 & 0.36309 & 0.29722 & 0.071 & +1  \\
29   & 0.24654 & -0.27031 & 0.01589 & -0.032 & +1  \\
30   & 0.27031 & 0.24654 & 0.01589 & -0.032 & +1  \\
31   & -0.27031 & -0.24654 & 0.01589 & -0.032 & +1  \\
32   & -0.24654 & 0.27031 & 0.01589 & -0.032 & +1  \\
33   & -0.77307 & 0.01265 & 0.01409 & 0.046 & -1  \\
34   & -0.01265 & -0.77307 & 0.01409 & 0.046 & -1  \\
35   & 0.01265 & 0.77307 & 0.01409 & 0.046 & -1  \\
36   & 0.77307 & -0.01265 & 0.01409 & 0.046 & -1  \\
37   & 0.04059 & 0.34601 & -0.28472 & 0.028 & +1  \\
38   & 0.34601 & 0.04059 & -0.28472 & 0.028 & +1  \\
39   & 0.34601 & 0.04059 & -0.28472 & 0.028 & +1  \\
40   & 0.04059 & 0.34601 & -0.28472 & 0.028 & +1
\end{tabularx}
    \end{minipage}
\end{table*}

\begin{table*}[!htb]
    \caption{Energies and Positions of Weyl nodes in PrAlSi.}
    \label{tab:WP_Si}
    \begin{minipage}{.5\linewidth}

      \centering
        \begin{tabularx}{0.97\textwidth}{p{0.6cm}|XXXXp{0.9cm}}
WP              & $k_x \ (\angstrom ^{-1})$ & $k_y \ (\angstrom ^{-1})$ & $k_z \ (\angstrom ^{-1})$ & $E$-$E_F$ (eV) & Charge  \\
\hline
\hline
1   & 0.37695 & 0.29462 & -0.00016 & 0.017 & -1  \\
2   & -0.29462 & 0.37695 & -0.00016 & 0.017 & -1  \\
3   & 0.29462 & -0.37695 & -0.00016 & 0.017 & -1  \\
4   & -0.37695 & -0.29462 & -0.00016 & 0.017 & -1  \\
5   & 0.0471 & -0.3534 & 0.2560 & -0.002 & -1  \\
6   & 0.3534 & 0.0471 & 0.2560 & -0.002 & -1  \\
7   & -0.3534 & -0.0471 & 0.2560 & -0.002 & -1  \\
8   & -0.0471 & 0.3534 & 0.2560 & -0.002 & -1  \\
9   & -0.2474 & -0.3046 & 0.0135 & 0.006 & -1  \\
10   & 0.3046 & -0.2474 & 0.0135 & 0.006 & -1  \\
11   & -0.3046 & 0.2474 & 0.0135 & 0.006 & -1  \\
12   & 0.2474 & 0.3046 & 0.0135 & 0.006 & -1  \\
13   & 0.76371 & 0.01571 & -0.01610 & 0.07 & +1  \\
14   & -0.01571 & 0.76371 & -0.01610 & 0.07 & +1  \\
15   & 0.01571 & -0.76371 & -0.01610 & 0.07 & +1  \\
16   & -0.76371 & -0.01571 & -0.01610 & 0.07 & +1  \\
17   & -0.03081 & 0.38176 & -0.25708 & 0.046 & -1  \\
18   & -0.38176 & -0.03081 & -0.25708 & 0.046 & -1  \\
19   & 0.38176 & 0.03081 & -0.25708 & 0.046 & -1  \\
20   & 0.03081 & -0.38176 & -0.25708 & 0.046 & -1
\end{tabularx}
    \end{minipage}%
    \begin{minipage}{.5\linewidth}
      \centering

       \begin{tabularx}{0.97\textwidth}{p{0.6cm}|XXXXp{0.9cm}}
WP              & $k_x \ (\angstrom ^{-1})$ & $k_y \ (\angstrom ^{-1})$ & $k_z \ (\angstrom ^{-1})$ & $E$-$E_F$ (eV) & Charge  \\
\hline
\hline
21   & -0.37618 & 0.29500 & 0.00003 & 0.018 & +1  \\
22   & -0.29500 & -0.37618 & 0.00003 & 0.018 & +1  \\
23   & 0.29500 & 0.37618 & 0.00003 & 0.018 & +1  \\
24   & 0.37618 & -0.29500 & 0.00003 & 0.018 & +1  \\
25   & -0.03095 & -0.38096 & 0.25678 & 0.047 & +1  \\
26   & 0.38096 & -0.03095 & 0.25678 & 0.047 & +1  \\
27   & -0.38096 & 0.03095 & 0.25678 & 0.047 & +1  \\
28   & 0.03095 & 0.38096 & 0.25678 & 0.047 & +1  \\
29   & 0.24850 & -0.30426 & 0.01882 & 0.004 & +1  \\
30   & 0.30426 & 0.24850 & 0.01882 & 0.004 & +1  \\
31   & -0.30426 & -0.24850 & 0.01882 & 0.004 & +1  \\
32   & -0.24850 & +0.30426 & 0.01882 & 0.004 & +1  \\
33   & -0.76389 & 0.01573 & 0.01658 & 0.07 & -1  \\
34   & -0.01573 & -0.76389 & 0.01658 & 0.07 & -1  \\
35   & 0.01573 & 0.76389 & 0.01658 & 0.07 & -1  \\
36   & 0.76389 & -0.01573 & 0.01658 & 0.07 & -1  \\
37   & 0.04701 & 0.35306 & -0.25580 & -0.002 & +1  \\
38   & -0.35306 & 0.04701 & -0.25580 & -0.002 & +1  \\
39   & 0.35306 & -0.04701 & -0.25580 & -0.002 & +1  \\
40   & -0.04701 & -0.35306 & -0.25580 & -0.002 & +1
\end{tabularx}
    \end{minipage}
\end{table*}

\nocite{*}
\bibliography{Yang_20oct2019}

\end{document}